\documentclass[preprint,review,12pt,number,sort&compress]{elsarticle_mod}

\pdfoutput=1

\usepackage{graphicx}
\usepackage{amssymb}
\usepackage{color}
\usepackage{subfig}

\begin{document}

\begin{frontmatter}

\title{
Structural properties of amorphous metal carbides; theory and experiment }

\author[Uppsala,Budapest]{Krisztina K\'adas\corref{@empty}}

\author[Uppchem]{Matilda Andersson}

\author[Valdivia]{Erik Holmstr\" om}

\author[Duisburg]{Heiko Wende}

\author[Uppmol]{Olof Karis}

\author[Uppchem]{Sigita Urbonaite}

\author[Uppmol]{Sergei M. Butorin}

\author[ESRF]{Sergey Nikitenko}

\author[ESRF]{Kristina O. Kvashnina}

\author[Uppchem]{Ulf Jansson}

\author[Uppsala]{Olle Eriksson\corref{cor1}}
\ead{Olle.Eriksson@physics.uu.se}

\cortext[cor1]{Corresponding author}

\address[Uppsala]{Division of Materials Theory, 
Department of Physics and Astronomy,
Uppsala University, Box 516, SE-751 20, Uppsala, Sweden}
\address[Budapest]{Institute for Solid State Physics and Optics, Wigner Research
Centre for Physics \\
H-1525 Budapest, P.O.Box 49, Hungary}
\address[Uppchem]{Department of Materials Chemistry,
Uppsala University, Box 538, 751 21 Uppsala, Sweden}
\address[Valdivia]{Instituto de F\'{\i}sica, Faculdad de Ciencias, 
Universidad Austral de Chile, Casilla 567, Valdivia, Chile}
\address[Duisburg]{Universit\" at Duisburg-Essen, Fakult\" at f\" ur Physik,
Experimentalphysik, Lotharstr. 1, D-47048 Duisburg, Germany}
\address[Uppmol]{Division of Molecular and Condensed Matter Physics,
Department of Physics and Astronomy, Box 516 751 20 Uppsala, Sweden}
\address[ESRF]{European Synchrotron Radiation Facility, BP 220, 6 rue Horowitz, 
38043 Grenoble, France}

\begin{abstract}
By means of theoretical modeling
and experimental synthesis and characterization,
we investigate the structural properties of 
amorphous Zr-Si-C.
Two chemical compositions are selected,
Zr$_{0.31}$Si$_{0.29}$C$_{0.40}$ and Zr$_{0.60}$Si$_{0.33}$C$_{0.07}$.
The amorphous structures are generated in the theoretical part of our work,
by the stochastic quenching (SQ) method,
and detailed comparison is made as regards structure and density of the
experimentally synthesized films.
These films are analyzed experimentally using X-ray absorption spectroscopy,  
transmission electron microscopy and X-ray diffraction.
Our results demonstrate
for the first time 
a remarkable agreement between theory and experiment
concerning
bond distances and atomic coordination of this complex amorphous metal carbide.
The demonstrated power of the SQ method opens up avenues for theoretical
predictions of amorphous materials in general.

\end{abstract}


\end{frontmatter}


\section{Introduction}
\label{intro}

Deposition of thin films from the vapor phase may lead to amorphous structures.  
A group of materials which often exhibit non-crystalline structure is various 
transition metal-metalloid films such as Cr-C, Zr-Si, and 
Fe-B.\cite{dehlinger2003,ziebert2011,wang2009}
Interestingly, they are also important components in 
metallic glasses formed during rapid quenching from a melt, 
suggesting that
the tendency to form amorphous structure is related to inherent properties 
of the elements and interactions between these elements. 
To obtain a more detailed understanding of the formation of amorphous 
structures and glasses we need to develop theoretical methods to model the
structure and bonding in these non-crystalline environments.  
This can be achieved by  
e.g. molecular dynamics (MD) \cite{thijssen99},
Monte Carlo simulation \cite{binder},
and reverse Monte Carlo simulation.\cite{rmc}
Unfortunately, these methods suffer from being computationally expensive
or to rely on interatomic potentials with sometimes questionable accuracy.
In this study we investigate another approach, namely the
stochastic quenching (SQ) method,\cite{holmstrom2009,holmstrom2010a} 
which combines
computational efficiency and accuracy in describing the chemical interaction.
In this method 
the atoms are placed randomly in the calculation cell, and then they are
relaxed using first-principles density functional (DFT) calculations
until the force on every atom is negligible.
It has been shown that this way of generating
amorphous structures is 
possible.\cite{holmstrom2009,holmstrom2010a} 
However, the present investigation is the first in which a detailed 
comparison between experiment and the SQ theory is made
in terms of structural properties, such as 
bond length and nearest neighbor coordination,
in particular for a material with complex chemical interactions.

In this study, we have selected the Zr-Si-C system as an amorphous model system,  
since transition metal carbides and silicides are known to have 
complex chemistry, with competing 
metallic and covalent bonds.\cite{gelatt83}
We deposit amorphous films using magnetron sputtering from elemental targets 
and confirm the structure experimentally with  several techniques. 
The choice of elements is based on several facts. 
First, Zr is a well-known base element in many metallic glasses. 
The atom has a large radius (1.59 \AA), favorable for amorphous structure 
formation when 
combined with several other elements with smaller radii.\cite{inoue2011}
Second, the Zr-Si system is well-known to produce amorphous thin films 
with magnetron sputtering for potential use as e.g. diffusion barriers.\cite{wang2009}
The addition of carbon to a ternary Zr-Si-C film will create a multicomponent system 
with different atomic radii favorable for glass formation.\cite{inoue2011}
Also, in a ternary Zr-Si-C film a wide variety of bond
types are formed
(metallic Zr-Zr, and covalent Zr-C, Zr-Si and Si-C bonds), 
which makes it possible to create a network 
structure which  should further favor an amorphous structure.   
The potential for applications
in Zr-Si-C has recently 
been discussed.\cite{krzanowski2006}

The paper is organized as follows. 
In Section \ref{theo}, we describe the theoretical methods used in this study, 
and summarize the most important details of the calculations. 
The experimental methods are described in Section \ref{exp}. 
The theoretical and experimental results are presented in Section \ref{res}.

\section{Theory}
\label{theo}

The amorphous structures were generated in the theory by the
stochastic quenching (SQ) method.
This method is designed
to describe amorphous structures in general,
although there is less experience with it
for complex materials with
competing natures of the chemical binding.
The amorphous structures are obtained by means of the following two steps of
the SQ method:
Firstly, the atoms are placed randomly in the calculation cell
under the constraint that no pair of atoms are closer than a small value
(typically 0.4 \AA). This constraint is required in order to avoid numerical
problems in the first few steps in the calculation.
Secondly, the
positions are relaxed
by means of a conjugate gradient method until the force on every atom is negligible.
We found that the self averaging of 150-200 atoms usually properly describes the
electronic properties at the thermodynamic limit.
\cite{holmstrom2010a,arhammar2011,Bock2010,amezaga2010,lizarraga2011,holmstrom2011}
In the present study we used 200 atoms for most of the calculations, and made
a few calculations with 400 atoms, for comparison.

The first-principles calculations were performed by
means of the projector augmented wave \cite{blochl94,kresse99} method as
implemented in the Vienna ab initio simulation package (VASP).
\cite{vasp1,vasp2,vasp3}.
This method is based on DFT.~\cite{hohenberg64,kohn65}
The exchange-correlation energy 
was calculated
using the generalized gradient approximation with the Perdew, Burke, and
Ernzerhof functional.\cite{pbe}
The calculations were considered converged when the potential energy difference
between atomic iterations was less than
10$^{-5}$ eV/atom,
and the forces on each atom were typically less than 0.005 eV/\AA.
A plane-wave energy cutoff of 400 eV was employed.
The calculations were performed using only the $\Gamma$ k-point.

\section{Experimental}
\label{exp}

The Zr-Si-C film was deposited by non-reactive, unbalanced,
DC magnetron sputtering in an ultra-high vacuum chamber
(base pressure of 1*10$^{-7}$ Pa).
Separate 2 inch elemental targets (Kurt J. Lesker Ltd) with a purity of
99.999\% for Si and C and 99.2-99.7\% (grade 702) for Zr were used.
The Ar plasma was generated at a constant pressure of 0.4 Pa with a
deposition temperature of approximately 350 $^{\circ}$C and bias of -50 V.
All depositions were made using a rotating substrate holder to ensure a
homogenous composition.
Prior to deposition the Si and SiO$_2$ substrates were cleaned using ultrasonic bath
in 2-propanol and ethanol for 5 minutes each and dried with nitrogen gas.
The deposition rate was 40-70 \AA/min
depending on composition
and the films for XRD,
XPS and resistivity analysis was deposited with a thickness of
$\sim$0.3 $\mu$m.
The chemical composition was analyzed by XPS using a
Physical Systems Quantum 2000 spectrometer with monochromatic Al $K\alpha$
radiation and using sensitivity factors determined from binary reference
samples of known composition.
The crystallinity of the film was studied using XRD and TEM.
The XRD analysis was made using a Siemens D5000 diffractometer and Cu $K\alpha$
radiation.
A gracing incidence (GI) scan with an incidence angle of 1$^{\circ}$
was used to increase the signal from the thin film in relation to the substrate.
For top view TEM the sample was prepared by depositing 40 nm thin film on a
NaCl substrate.
The NaCl was then dissolved in distilled water after deposition releasing
the deposited film, which then was put on a carbon-coated copper grid and
analyzed using a JEOL JEM-2100 operated at 200 kV.
For the resistivity measurements
each sample was measured six times and mean value and 95\% confidence
interval was calculated.

The XAS measurements were performed at BM26A (Dutch-Belgian beamline "DUBBLE")
\cite{nikitenko08}
of the European Synchrotron Radiation Facility (ESRF) in Grenoble, France.
The incident energy was selected using the $<$111$>$ reflection from a double
Si crystal monochromator with bendable 2nd crystal for sagital (horizontal)
focusing. Vertical focusing and rejection of higher harmonics was achieved with Si
mirror with Pt layer at an angle of 1.8 mrad relative to the incident beam.
The incident X-ray beam had a flux of approximately
1 x 10$^{11}$ photons/s on the sample position.
The XAS data were measured in fluorescence mode by 9-element monolithic
Ge detector \cite{derbyshire99}
at Zr K edge (17988 eV) at room temperature.
The Zr-Si-C thin films on SiO$_2$ substrates
were conventionally mounted on a thin tape.
The energy calibration was performed on Zr foil.
The data were recorded with a constant 0.05 \AA $^{-1}$ step in $k$-space,
in the energy range 17.75-18.85 keV.
The intensity was normalised to the incident flux.
The total energy resolution was estimated to be $\sim$3.5 eV.

Spectra were analysed using the \textsc{iXAFS} implementation of
\textsc{IFEFFIT}, \textsc{Athena}, and \textsc{Artemis}.
\cite{ravel2005,newville2001}
Model data were obtained from averages of many calculations of EXAFS
data obtained from local structures present in
the description of the amorphous material,
using the \textsc{FEFF} software.
\cite{zabinsky95,rehr92,mustre91}

\section{Results and Discussion}
\label{res}

\begin{figure}
\begin{center}
\includegraphics[width=14cm]{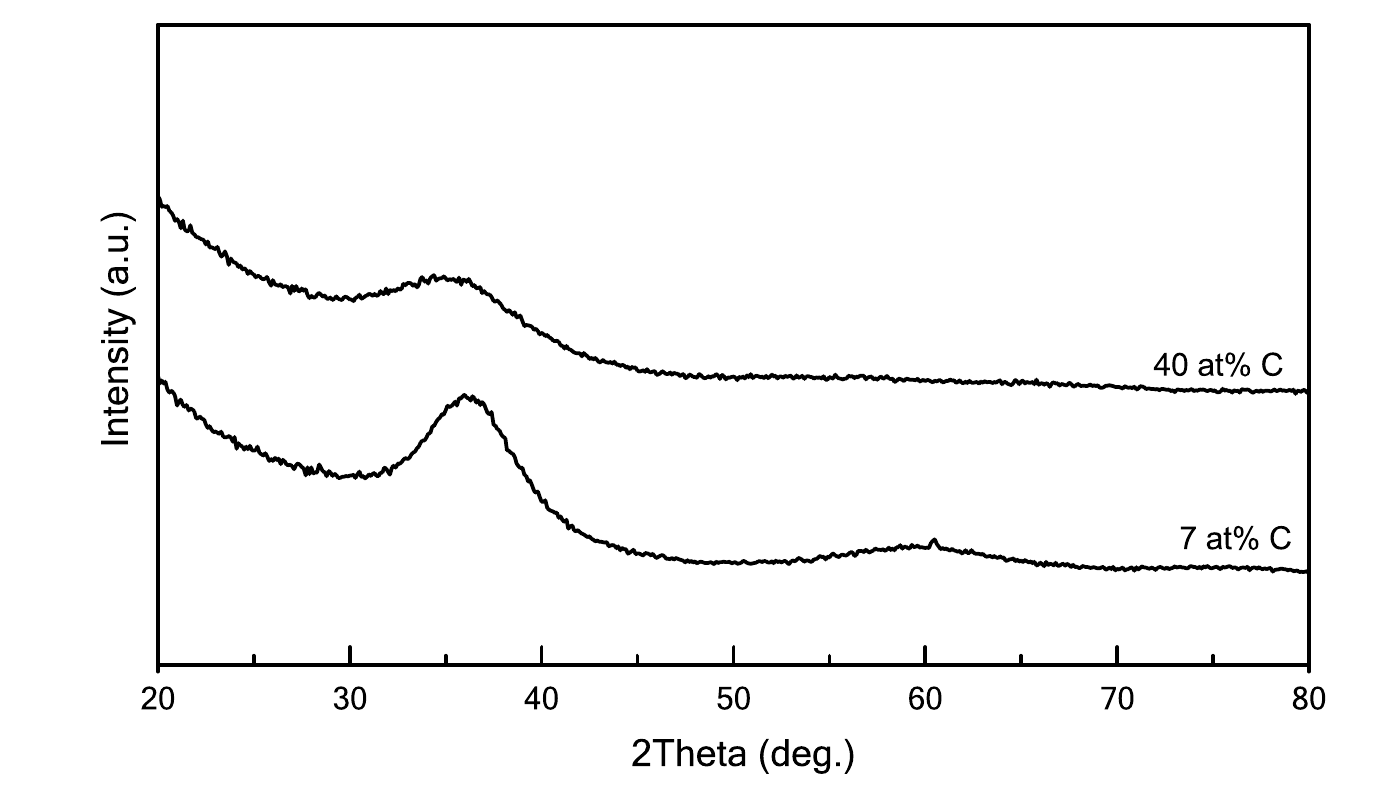}
\caption{}
\label{fig_diffr}
\end{center}
\end{figure}

\begin{figure}
\begin{center}
\includegraphics[width=13cm]{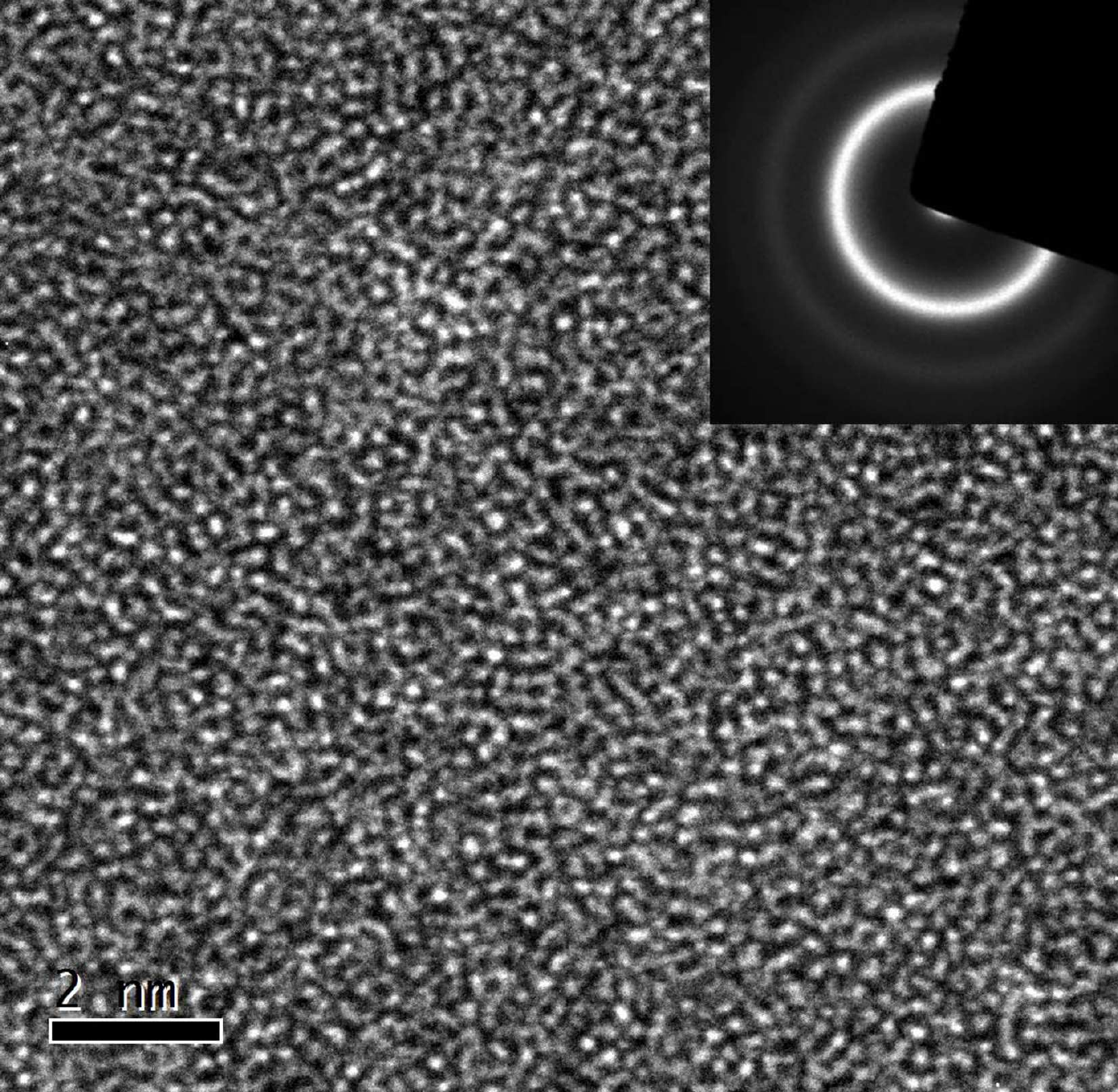}
\caption{}
\label{fig_tem}
\end{center}
\end{figure}

On the experimental side, amorphous structures are typically generated by 
fast cooling from melts 
or with quenching of atoms
during growth
directly from the gas phase.
Two films with composition 
Zr$_{0.60}$Si$_{0.33}$C$_{0.07}$ and
Zr$_{0.31}$Si$_{0.29}$C$_{0.40}$ 
(according to our XPS analysis) were here deposited using magnetron sputtering. 
In our XRD measurements
both films exhibited identical types of structure.
Figure \ref{fig_diffr} shows the GI-XRD diffractograms of the 
Zr$_{0.60}$Si$_{0.33}$C$_{0.07}$ and
Zr$_{0.31}$Si$_{0.29}$C$_{0.40}$
films. 
Two broad features, 
one at 36$^{\circ}$ and the other one at 60$^{\circ}$
is observed in the diffractogram.
Another small peak is observed at 61$^{\circ}$
for the Zr$_{0.60}$Si$_{0.33}$C$_{0.07}$ film, which is only a contribution 
from the Si(111) substrate.
This type of diffractograms with only broad features 
is characteristic for an X-ray amorphous material, 
where these features are a consequence of the close-range order in the material.
Fig. \ref{fig_tem} shows the TEM image of the Zr$_{0.60}$Si$_{0.33}$C$_{0.07}$ film. 
No indication of crystallinity is visible and the film 
shows a completely amorphous structure. 
Also, the energy filtered electron diffraction (inset) shows only broad, 
featureless rings without indication of any crystalline contribution. 
TEM and XRD together confirm that both films 
are amorphous.

\begin{figure}
\begin{center}
\includegraphics[width=14cm]{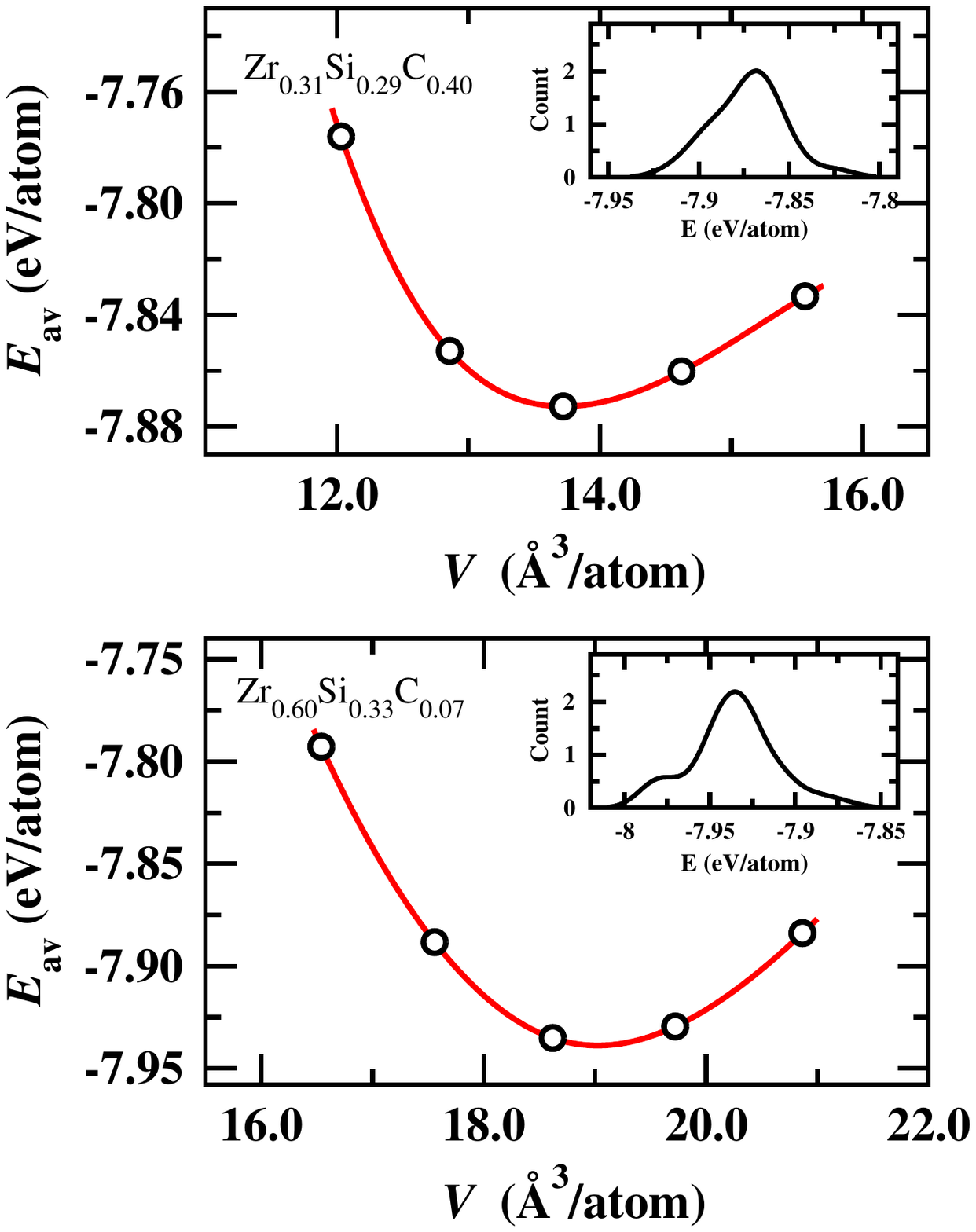}
\caption{}
\label{fig_energ}
\end{center}
\end{figure}

\begin{figure}
\begin{center}
\hspace{-1cm}
\includegraphics[width=17cm]{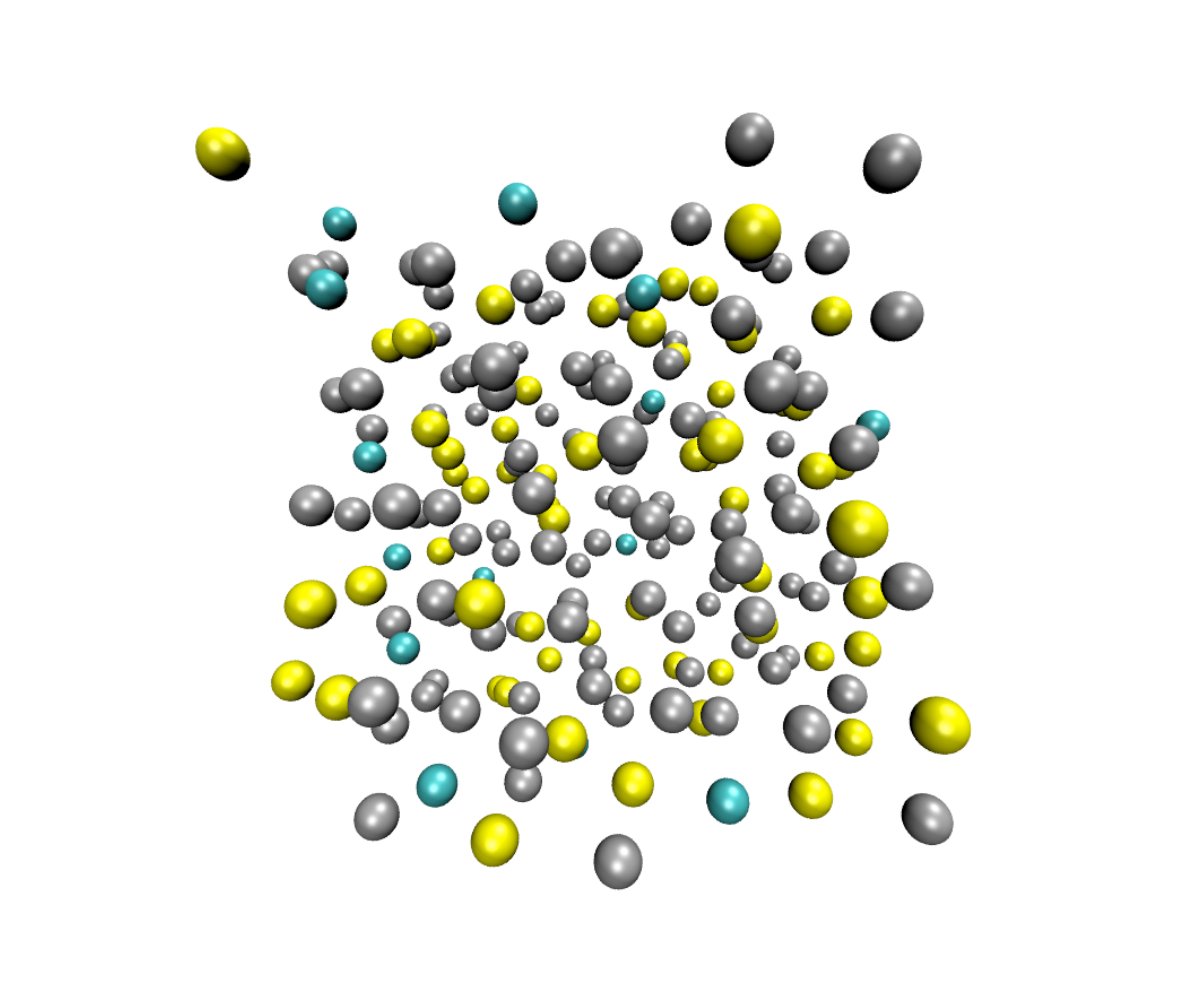}
\caption{}
\label{fig_str}
\end{center}
\end{figure}

The amorphous structures were generated in the theory by the SQ
method.
We calculated the energies of 50 stochastic structures
of both compositions studied in the experiments,
at different volumes,
with supercells of 200 atoms (Fig. \ref{fig_energ}).
We found an amorphous structure in all generated configurations.
No partial crystallization or porosity was observed 
(Fig. \ref{fig_str}).
The calculated energies at constant $V$ show roughly a Gaussian distribution 
for both compositions (insets in Fig. \ref{fig_energ}).
We have determined the equilibrium volume ($V_{\rm 0}$)
and bulk modulus ($B_{\rm 0}$)
by fitting an exponential Morse-type function to the average energies
of Fig. \ref{fig_energ},
and obtained $V_{\rm 0}$=13.72 \AA$^3$/atom for
Zr$_{0.31}$Si$_{0.29}$C$_{0.40}$, 
and $V_{\rm 0}$=19.08 \AA$^3$/atom  for
Zr$_{0.60}$Si$_{0.33}$C$_{0.07}$. 
Accordingly, our calculated equilibrium densities are
$\varrho_{\rm 0}$=4.99 g/cm$^3$ for Zr$_{0.31}$Si$_{0.29}$C$_{0.40}$, and
$\varrho_{\rm 0}$=5.64 g/cm$^3$ for Zr$_{0.60}$Si$_{0.33}$C$_{0.07}$.
This is in line with the experimental densities of the corresponding
binary crystalline phases:
ZrC (6.730 g/cm$^3$ \cite{martienssen05}),
ZrSi$_2$ (4.883 g/cm$^3$ \cite{zatorska2002}),
and Zr$_5$Si$_3$ (5.998 g/cm$^3$ \cite{celis91}).
We calculate 
$B_{\rm 0}$=86 GPa for 
Zr$_{0.31}$Si$_{0.29}$C$_{0.40}$, and $B_{\rm 0}$=116 GPa for
Zr$_{0.60}$Si$_{0.33}$C$_{0.07}$.
Within accuracy,
$B_{\rm 0}$ in Zr$_{0.60}$Si$_{0.33}$C$_{0.07}$ is about the same as that in
crystalline ZrSi$_2$ ($B_{\rm ZrSi_2}$=114.8 GPa \cite{peun95}), 
$B_{\rm 0}$ in Zr$_{0.31}$Si$_{0.29}$C$_{0.40}$, however, is noticeably lower than 
that.
$B_{\rm 0}$ of both amorphous materials is smaller than that of
crystalline Zr$_5$Si$_3$ ($B_{\rm Zr_5Si_3}$=152 GPa \cite{celis91}).
As expected, the theoretical $B$ of amorphous Zr-Si-C 
is significantly smaller than that of the crystalline carbides
($B_{\rm ZrC}$=223 GPa, $B_{\rm SiC}$=211 GPa).

We note that in amorphous materials, in contrast to crystalline materials,
not only bond lengths can change upon compression,
but also bond angles may vary, even to the extent that atoms may rearrange
under compression.
This invariably reduces the curvature of the energy versus volume relationship,
when compared to a crystalline material with a reduced bulk modulus as the result.
It is possible that this explains the much lower bulk modulus for amorphous
Zr$_{0.31}$Si$_{0.29}$C$_{0.40}$ 
($B_{\rm 0}$=86 GPa)
when compared to ZrC.
The presence of Si also changes the bonding situation which may contribute to
the relatively low bulk modulus of Zr$_{0.31}$Si$_{0.29}$C$_{0.40}$.
It should also be mentioned here that the larger bulk modulus of
Zr$_{0.60}$Si$_{0.33}$C$_{0.07}$ ($B_{\rm 0}$=116 GPa) compared to that of
Zr$_{0.31}$Si$_{0.29}$C$_{0.40}$ is somewhat surprising.
The compound with more carbon has a higher density, and is expected to have more
of strong Si-C bonds, which is expected to increase the bulk modulus,
compared to a material with lower C content.
Our calculations give the opposite trend.
As we show in our study, the structures generated by the SQ method agree very well
with observations.
Hence it remains to be seen if the calculated trend in bulk modulus is a true
effect, or possibly the result of the numerical treatment of the 50 configurations
considered at each volume.

\begin{figure}
\begin{center}
\includegraphics[width=17cm]{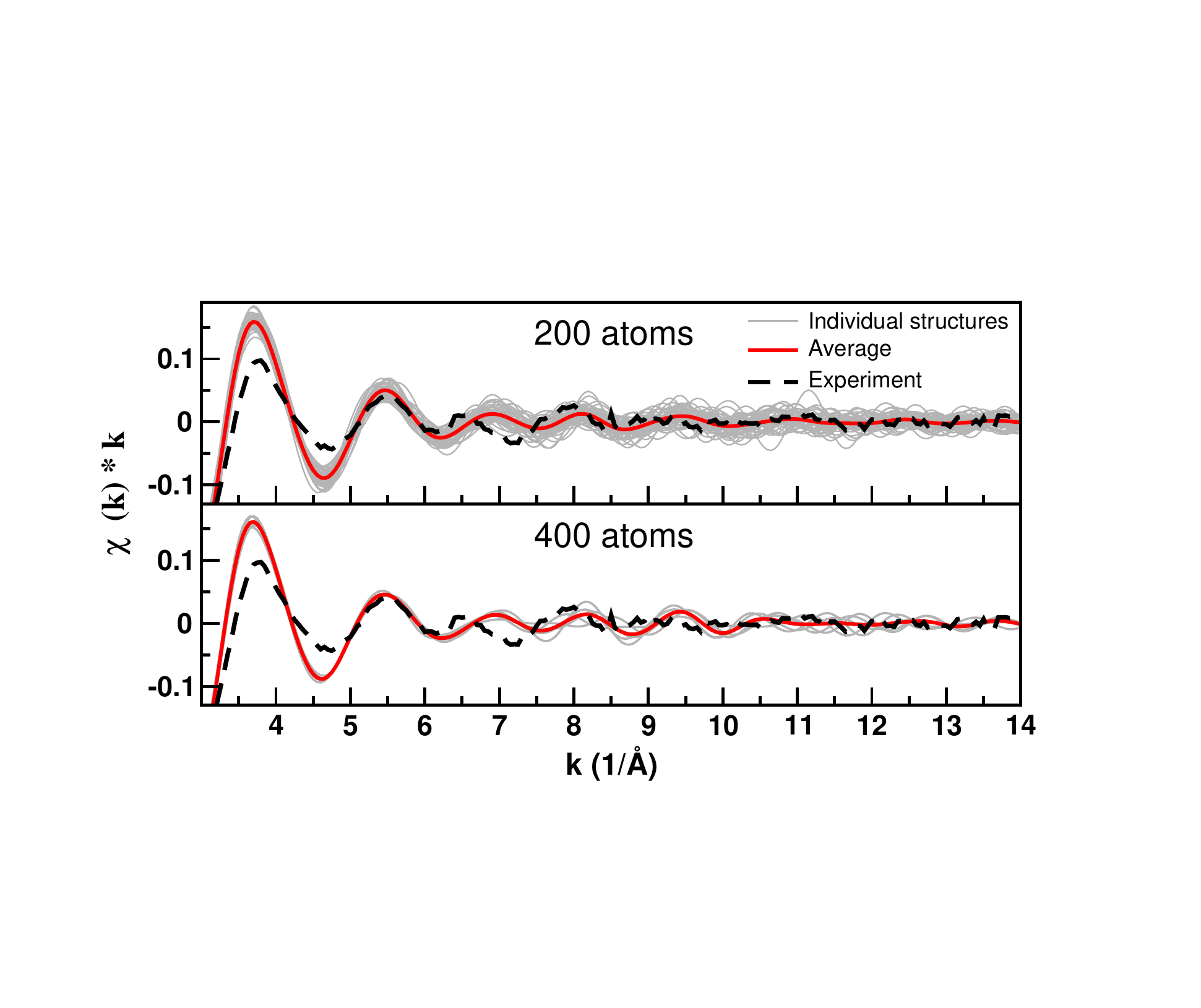}
\caption{}
\label{fig_feff31}
\end{center}
\end{figure}

\begin{figure}
\begin{center}
\includegraphics[width=17cm]{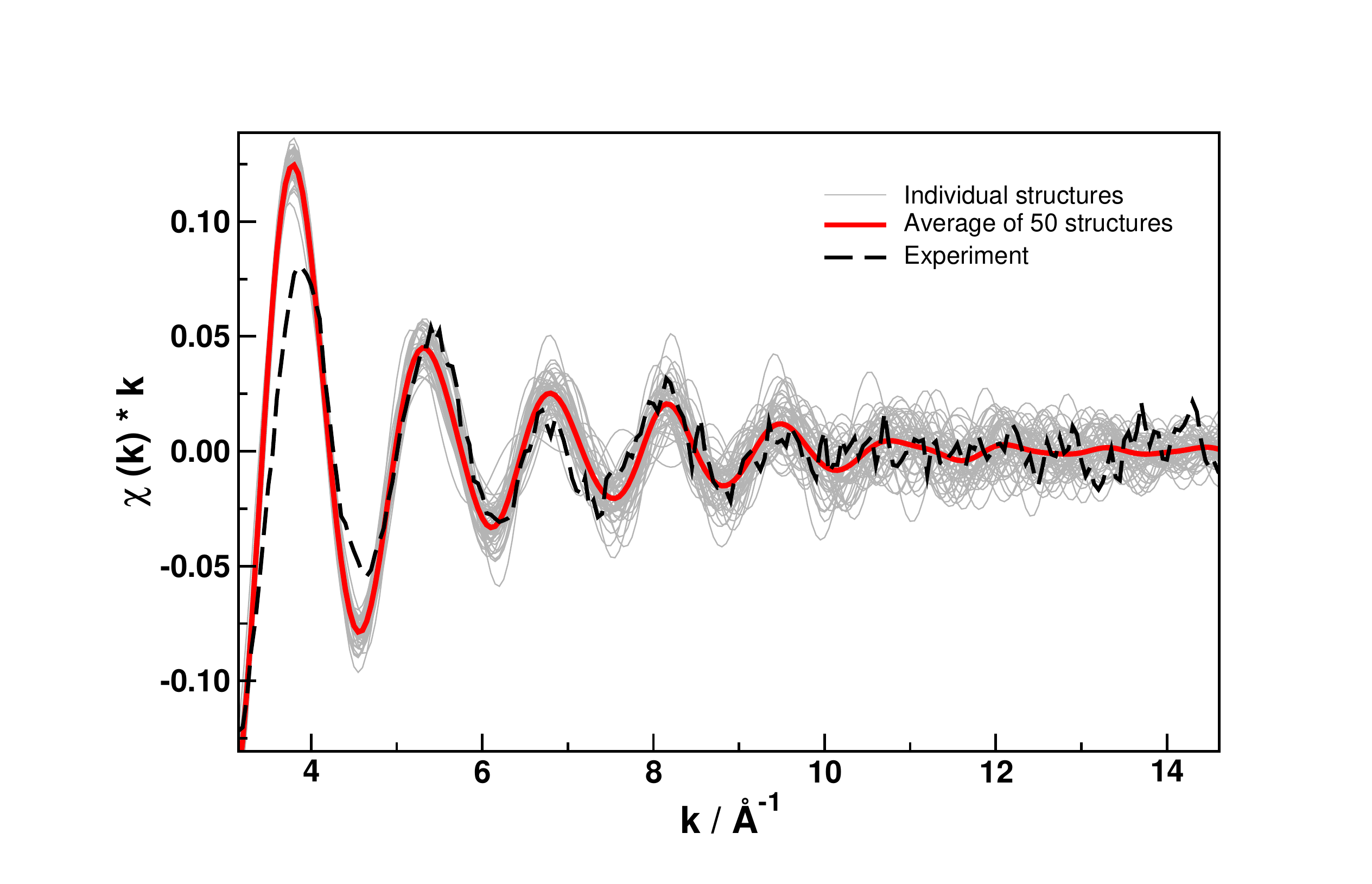}
\caption{}
\label{fig_feff60}
\end{center}
\end{figure}

In order to compare the structural properties of the theoretical and
experimental amorphous phases,
in Fig. \ref{fig_feff31} and \ref{fig_feff60}
we compare the simulated 
X-ray absorption fine structure (EXAFS at the Zr K-edge) signals of
Zr$_{0.31}$Si$_{0.29}$C$_{0.40}$ and Zr$_{0.60}$Si$_{0.33}$C$_{0.07}$
to the experimental results.
The simulations of the EXAFS spectra were performed by means of real space
multiple-scattering theory as implemented in the FEFF code \cite{ankudinov98}.
The size of our cells prohibited the use of self-consistent potentials for the
multiple scattering calculation of the X-ray absorption coefficients so we
distributed overlapping atomic potentials on the sites in our amorphous 200 and
400 atom cells. The multiple scattering path expansion was used as the full
multiple scattering approach loses accuracy at high k-values.
In order to simulate periodic boundary conditions, we constructed a 3x3x3
supercell and calculated the normalized EXAFS signal $\chi^C_i(\rm k)$ from
each Zr atom in the
center cell. The index $i$ denotes individual Zr atoms, and the index $C$
denotes individual 200 atom or 400 atom cells obtained by means of the SQ technique.

For each cell $C$, the calculated
normalized EXAFS spectra from each individual Zr atom $\chi^C_i(\rm k)$ were 
averaged to form the final $\chi^C(\rm k)$.
The theoretical spectra were calculated at 0 K and at room temperature 
by means of the
correlated Debye model. The Debye temperature for the amorphous material is
unknown, so we estimated a value of 600 K for the calculations
that was close to the Debye temperatures of 
ZrC (614 K - 680 K \cite{houska64,lawson2008}), 
ZrSi$_2$ (495 K \cite{mcrae84}),
and Zr$_5$Si$_3$ (480 K \cite{celis91}).

The theoretical curves of the 50 stochastic structures with 200 atoms 
(grey lines in Fig. \ref{fig_feff31} and \ref{fig_feff60})
are very close to each other, especially in the small $k$ region,
and with increasing $k$ they become more diffuse.
We find that 
the overall agreement between
the averages of the 50 individual structures
(red lines) and the experimental signals
(dashed black line) is good. 
In particular, we obtain excellent agreement between theory and experiment 
in the small $k$ region.
When comparing Fig. \ref{fig_feff31} and Fig. \ref{fig_feff60}
we note that Fig. \ref{fig_feff60} reproduces observations almost perfectly,
whereas a disagreement may be seen in Fig. \ref{fig_feff31}
in the region of 6.5-7  \AA$^{-1}$.
We made a few additional calculations with 400 atoms per unit cell.
As expected, due to the better statistics, the curves of the individual stochastic 
structures are closer to each other than for the 200 atom calculations
[Fig. \ref{fig_feff31}].
The average of the individual structures practically does not change compared
to the 200 atom results, 
which clearly shows that a 200 atom unit cell
is sufficient to describe amorphous Zr-Si-C.

Returning to the disagreement between theory and experiment in the region of
6.5-7 \AA$^{-1}$, for the sample with low Zr concentration, 
it can be analyzed further by making a Fourier
transformation of the data in Fig. \ref{fig_feff31}. 
This transform (data not shown) shows that 
two bond lengths at distance just above 2.5 \AA \ are located at 0.2 \AA \ shorter 
distances in the theoretical data compared to the experiment. 
Although this may be viewed as a smaller error to an overall rather satisfactory 
agreement between experiment and theory, it can most likely be traced back to 
the very fast quenching of atomic coordinates in the SQ method, which is a 
somewhat slower process in the experimental samples. 
This may be realized by noting that for the low Zr concentration case, 
the two peaks for which the SQ method makes a 10 \% error 
are associated to the average Zr-Zr and Zr-Si bond lengths.
In this sample we expect few nearest neighbor Zr-Zr coordination, and a reduced 
amount of Zr-Si bonds, due to a preferred Zr-C bonding in combination with a rather 
large amount of C in the sample. However, an inspection of the simulated geometries 
gives that the SQ method does find a finite amount of Zr-Zr nearest 
neighbor coordination, due to the rapid quenching. 
For the sample with larger Zr concentration the amount of C is very low, and hence 
neither simulations nor experiment can exclude a nearest neighbor Zr-Zr 
coordination, resulting in a better agreement between theory and experiment for the 
EXAFS data.

\begin{figure}
\begin{center}
\includegraphics[width=14cm]{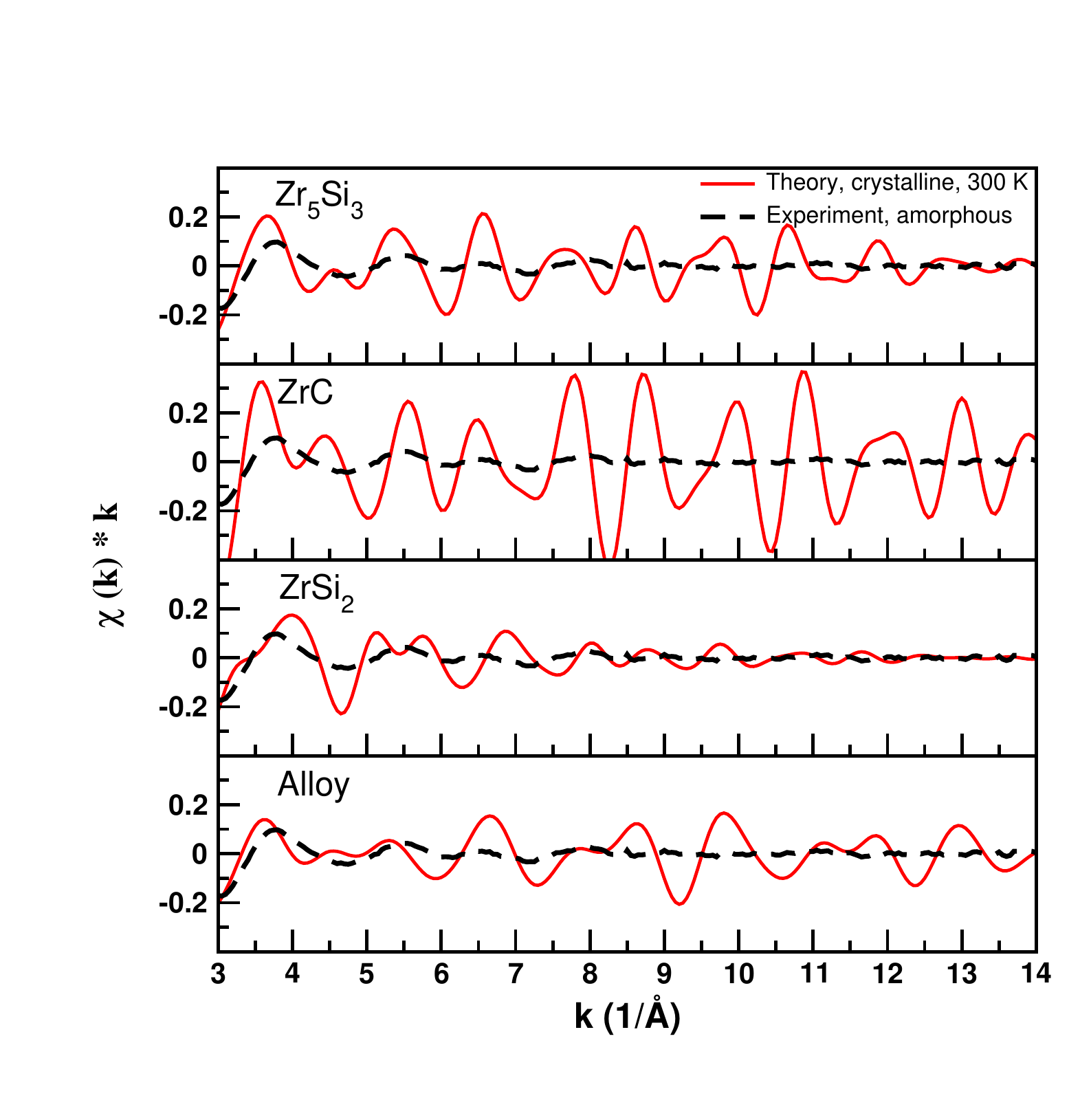}
\caption{}
\label{fig_feffcryst}
\end{center}
\end{figure}

To demonstrate that the agreement between theory and experiment is really due to 
an amorphous structure, 
we also calculated the EXAFS spectra of some known crystalline compounds of Zr, Si,
and C
(Fig. \ref{fig_feffcryst}).
In this EXAFS calculation we used the experimental crystal structures.
The Debye temperatures used for these crystals were 680 K (ZrC)
\cite{lawson2008}, 495 K (ZrSi$_2$) \cite{mcrae84}, 480 K (Zr$_5$Si$_3$)
\cite{celis91}, and 600 K (Random alloy)\cite{footnote1}.
The simulated EXAFS curves of these compunds do not agree at all 
with experiment.
We also examined a case, where the
calculated composition (Zr$_{0.40}$Si$_{0.20}$C$_{0.40}$)
differs slightly from the experimental one (Zr$_{0.43}$Si$_{0.30}$C$_{0.27}$)
to a small extent.
For such a composition, the simulated curve does not
agree with the experimental one, neither in the small, nor in the large
$k$ region (data not shown).
This shows that both accurate structural information as well as
precise chemical composition is needed to reproduce the experimental EXAFS data.
The data in Figs. \ref{fig_feff31} and \ref{fig_feff60}
are the first direct comparison on a structural level, between the SQ method 
and observations, and the agreement shows that 
a computationally efficient theory provides accurate 
atomic coordinates for a complex amorphous metal carbide.

\begin{figure}
\begin{center}
\includegraphics[width=17cm]{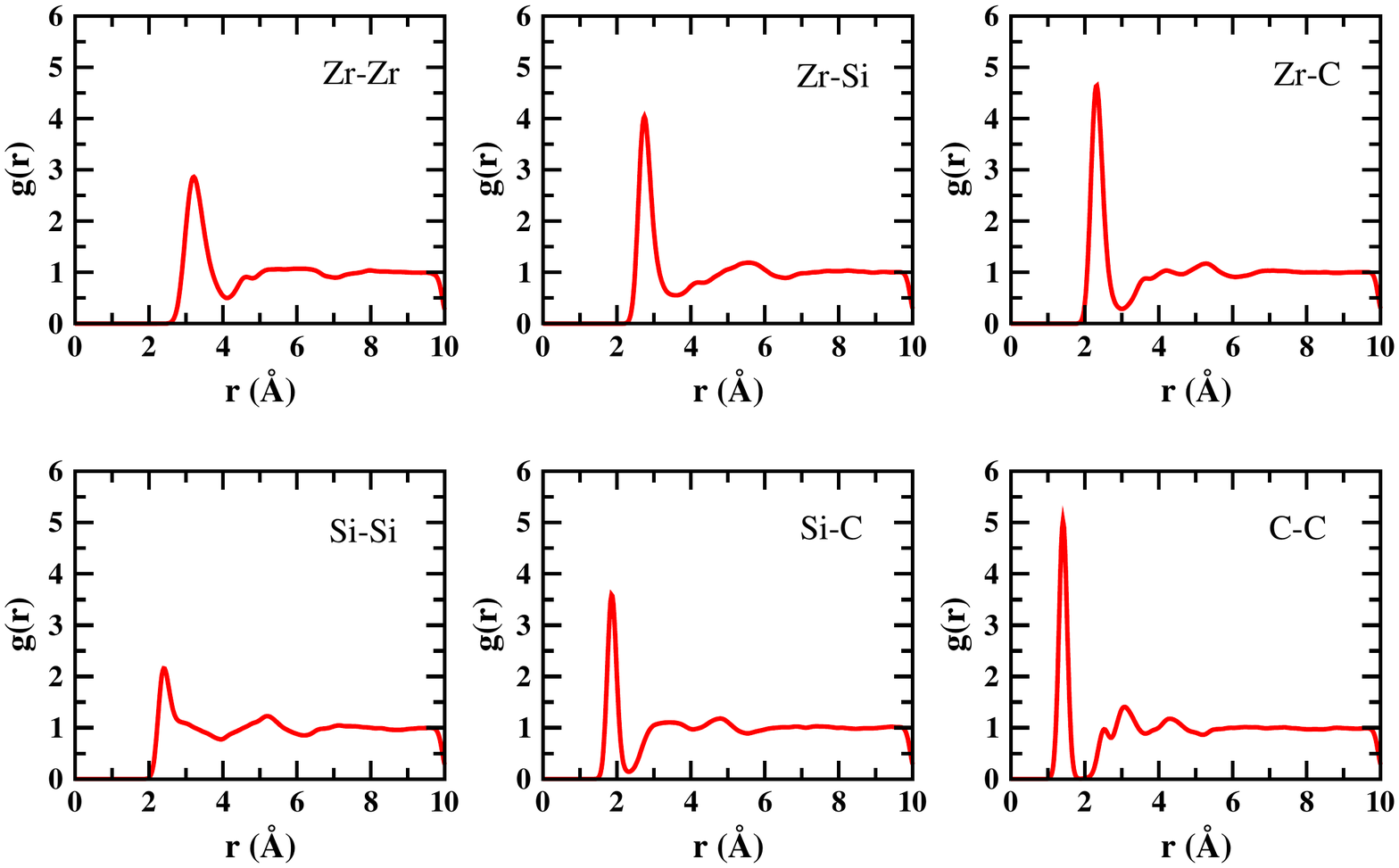}
\caption{}
\label{fig_rdf31}
\end{center}
\end{figure}

\begin{figure}
\begin{center}
\includegraphics[width=17cm]{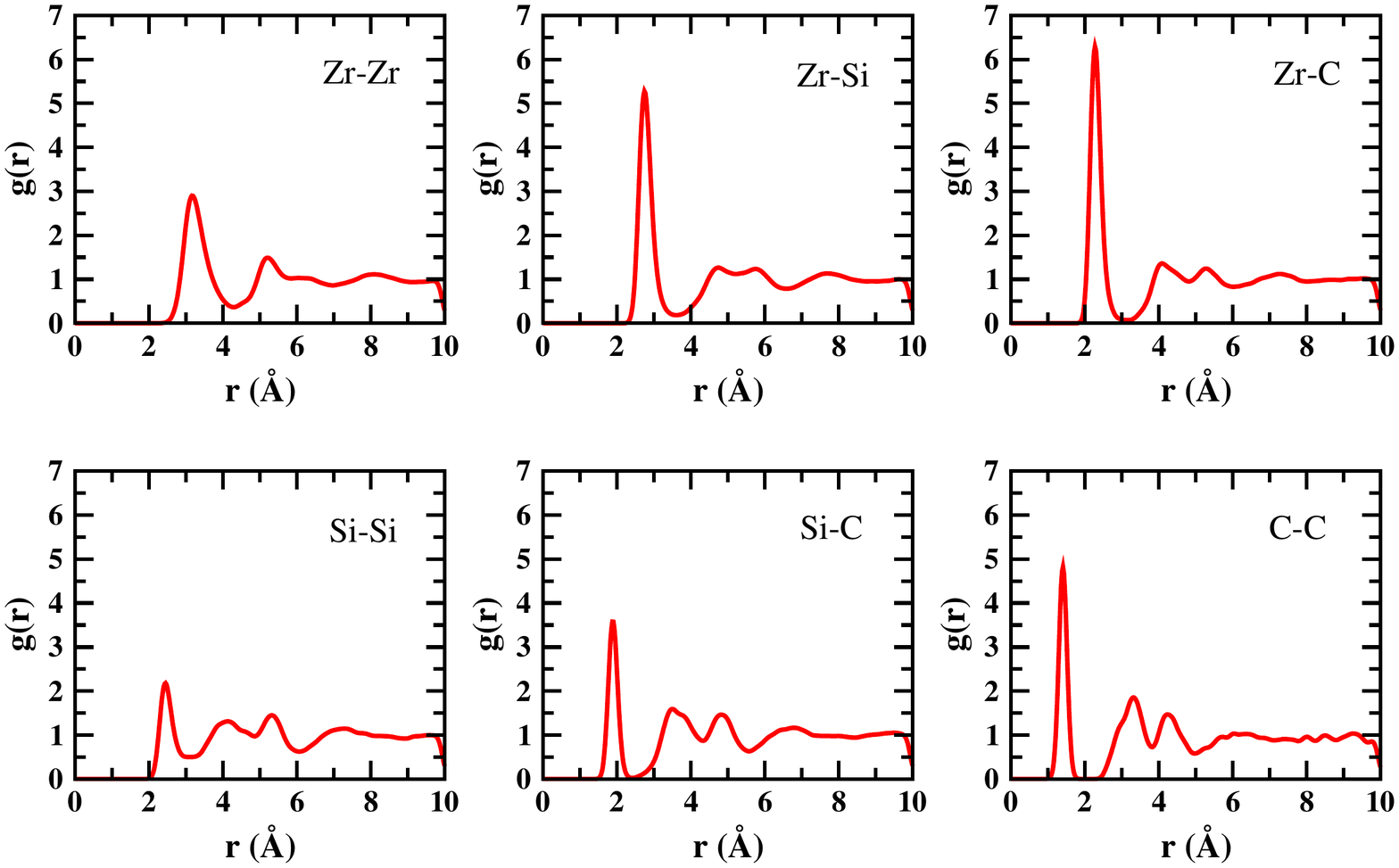}
\caption{}
\label{fig_rdf60}
\end{center}
\end{figure}

In the following, we present additional structural details about amorphous 
Zr-Si-C.
The theoretical average partial radial distribution functions (RDF)
calculated for amorphous Zr$_{0.31}$Si$_{0.29}$C$_{0.40}$ and 
Zr$_{0.60}$Si$_{0.33}$C$_{0.07}$
(Figs. \ref{fig_rdf31} and \ref{fig_rdf60})
show short range order up to 6 \AA \ for the
C bonds (Zr-C, Si-C, and C-C), 
and up to 7-8 \AA \ for the longer Zr-Zr, Zr-Si, and Si-Si bonds.
In Table \ref{tab2} the calculated average nearest neighbor distances
are compared to distances in typical crystalline phases.
Atoms are considered
nearest neighbors (and also counted in the coordination numbers later),
if they are closer than the first minimun in the corresponding partial RDF curves.
We used the following cutoff distances
for Zr$_{0.31}$Si$_{0.29}$C$_{0.40}$ (and Zr$_{0.60}$Si$_{0.33}$C$_{0.07}$):
$d_{\rm c,Zr-Zr}$=4.11 (4.29) \AA,
$d_{\rm c,Zr-Si}$=3.60 (3.61) \AA,
$d_{\rm c,Zr-C}$=3.00 (3.15) \AA,
$d_{\rm c,Si-Si}$=3.52 (3.10) \AA,
$d_{\rm c,Si-C}$=2.33 (2.41) \AA, and
$d_{\rm c,C-C}$=1.89 (1.92) \AA.
For most pairs the distances 
are comparable to the corresponding crystalline bonds.
The pairs with shorter bonds correlate with a reduced coordination of carbon.
A lowered coordination means that the remaining bonds are stronger than
in the crystalline phase.
Such bond-shortening mechanism was observed in MAX phases.\cite{palmquist04}

\begin{figure}
\subfloat[]{\label{fig_zr}\includegraphics[width=0.4\textwidth]{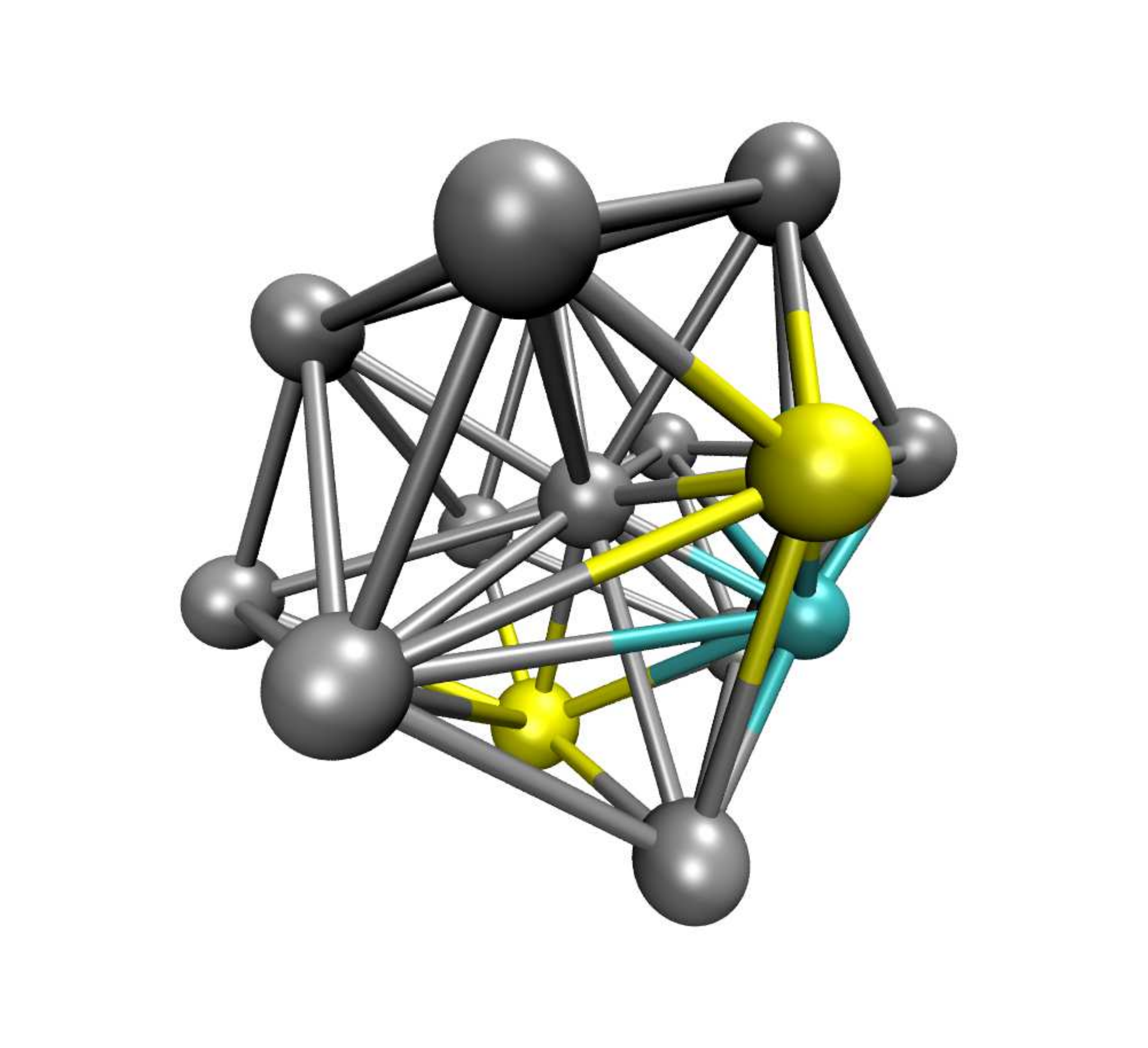}}
\subfloat[]{\label{fig_si}\includegraphics[width=0.4\textwidth]{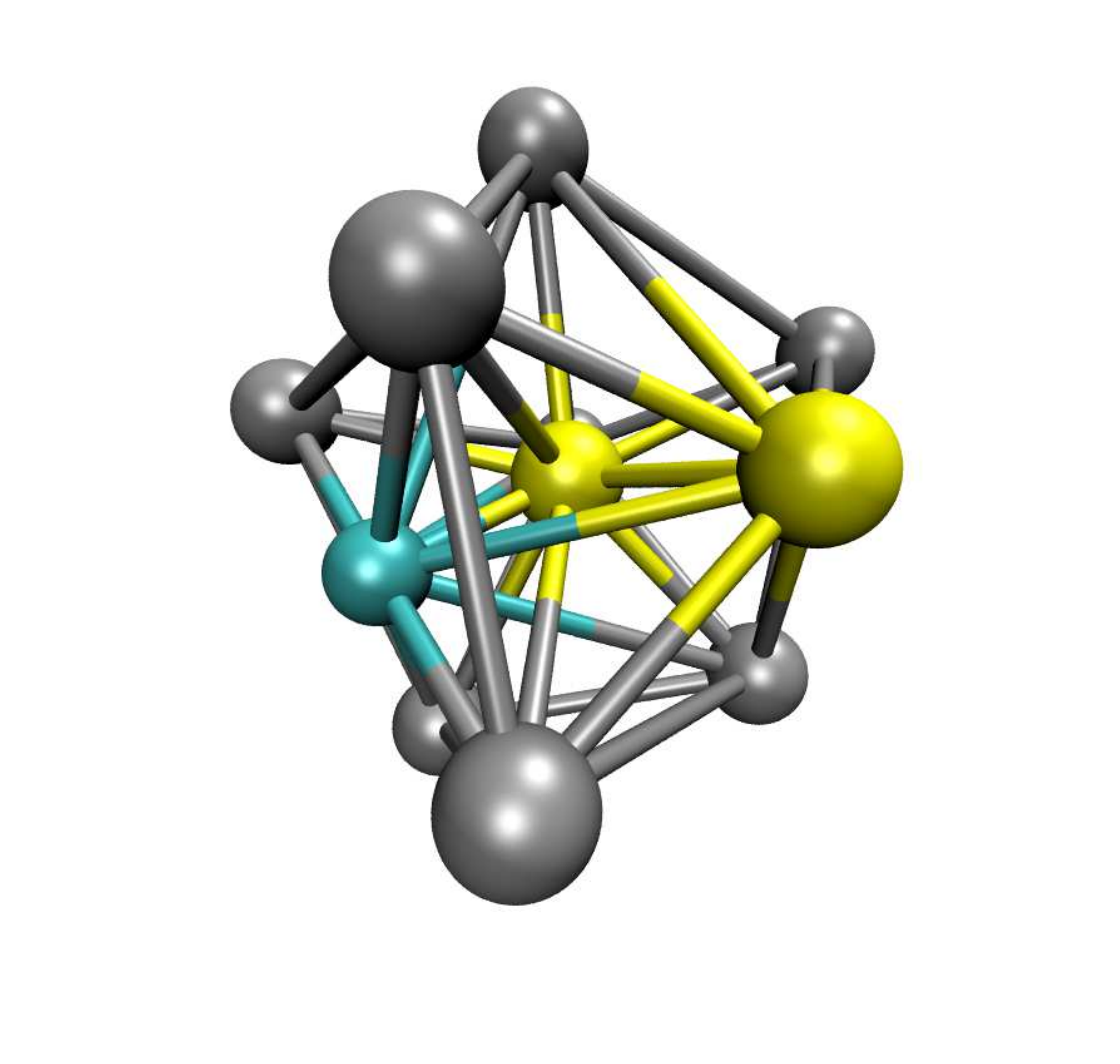}}
\subfloat[]{\label{fig_c}\includegraphics[width=0.4\textwidth]{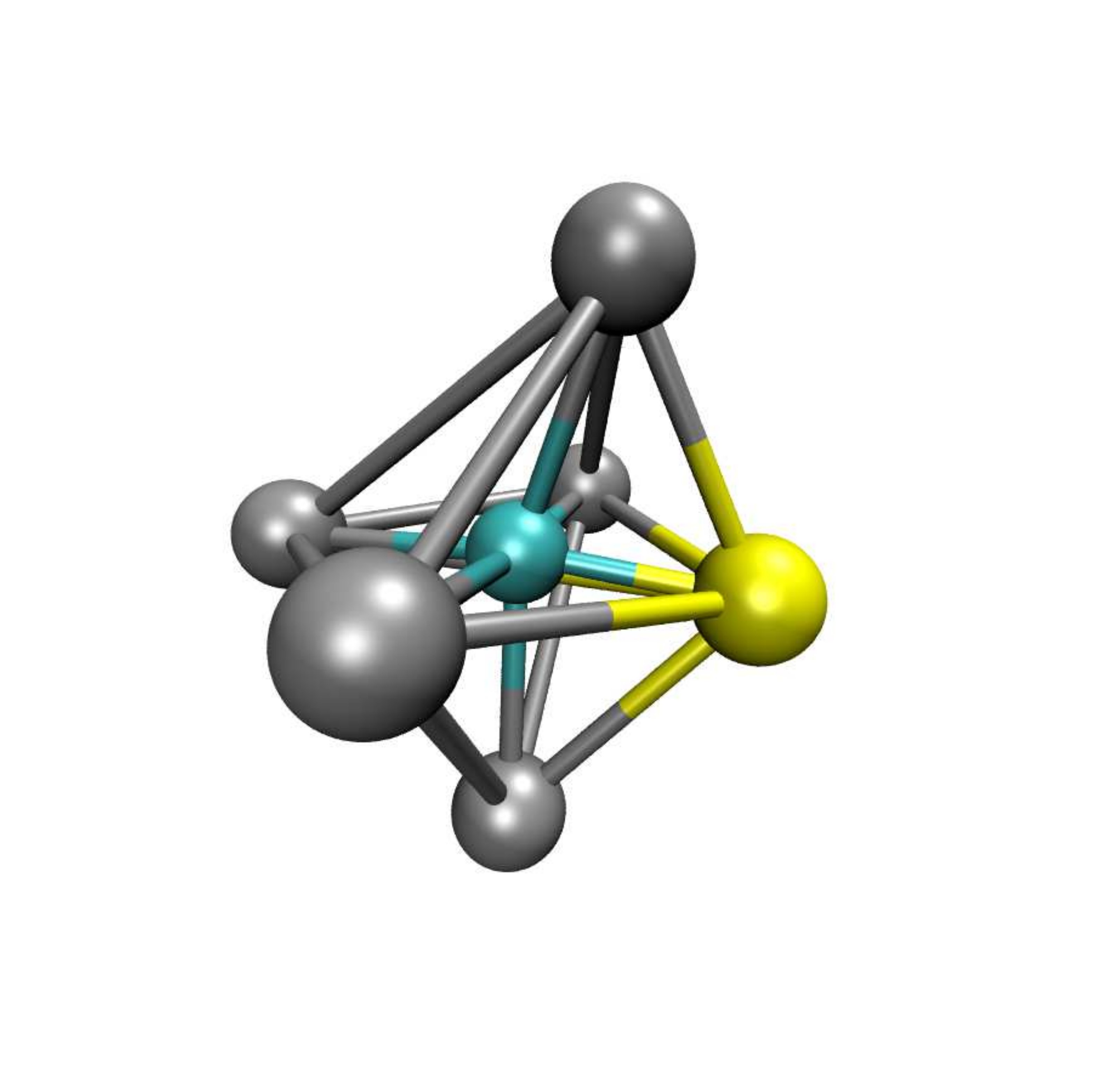}}
\caption{}
\label{fig_local}
\end{figure}

On the average,
Zr atoms are coordinated by 15.4 atoms in Zr$_{0.31}$Si$_{0.29}$C$_{0.40}$,
and by 14.9 atoms in Zr$_{0.60}$Si$_{0.33}$C$_{0.07}$ (Table \ref{tab3}).
Both are larger than the coordination number in 
hexagonal close-packed (hcp) structure ($n_{\rm hcp}$=12).
Such high coordination is not unusual in amorphous structures.\cite{egami2003}
Zr atoms are mostly coordinated by Zr atoms (Table \ref{tab3} and Fig. \ref{fig_zr}),
which may be primarily due to their high concentration.
Si atoms 
have a slight preference to have Zr neighbours (Fig. \ref{fig_si}),
which is pronounced in Zr$_{0.60}$Si$_{0.33}$C$_{0.07}$, where
the Zr-Si coordination is more than double that of 
the Si-Si coordination (Table \ref{tab3}),
and 
Zr has almost ten times larger number of Si nearest neighbors than Si
(see Table \ref{tab4}).
Due to the small C content of amorphous Zr$_{0.60}$Si$_{0.33}$C$_{0.07}$,
$n_{\rm Zr,C}$, $n_{\rm Si,C}$ and $n_{\rm C,C}$ 
(defined in caption of Table \ref{tab3}) 
are well below one,
and C atoms are mostly surrounded by Zr atoms
(Fig.  \ref{fig_c}).
As expected, the number of C-C, Si-C and Zr-C bonds is significantly larger
in Zr$_{0.31}$Si$_{0.29}$C$_{0.40}$ 
(Table \ref{tab4}).
Here C atoms are mostly coordinated by Zr atoms, and on the average, each
of them has at least one C nearest neighbors.
Figure \ref{fig_local} shows that the structure can be described as a 
network structure of metal-metal, metal-carbide and metal-silicon bonds.

\begin{figure}
\includegraphics[width=14cm]{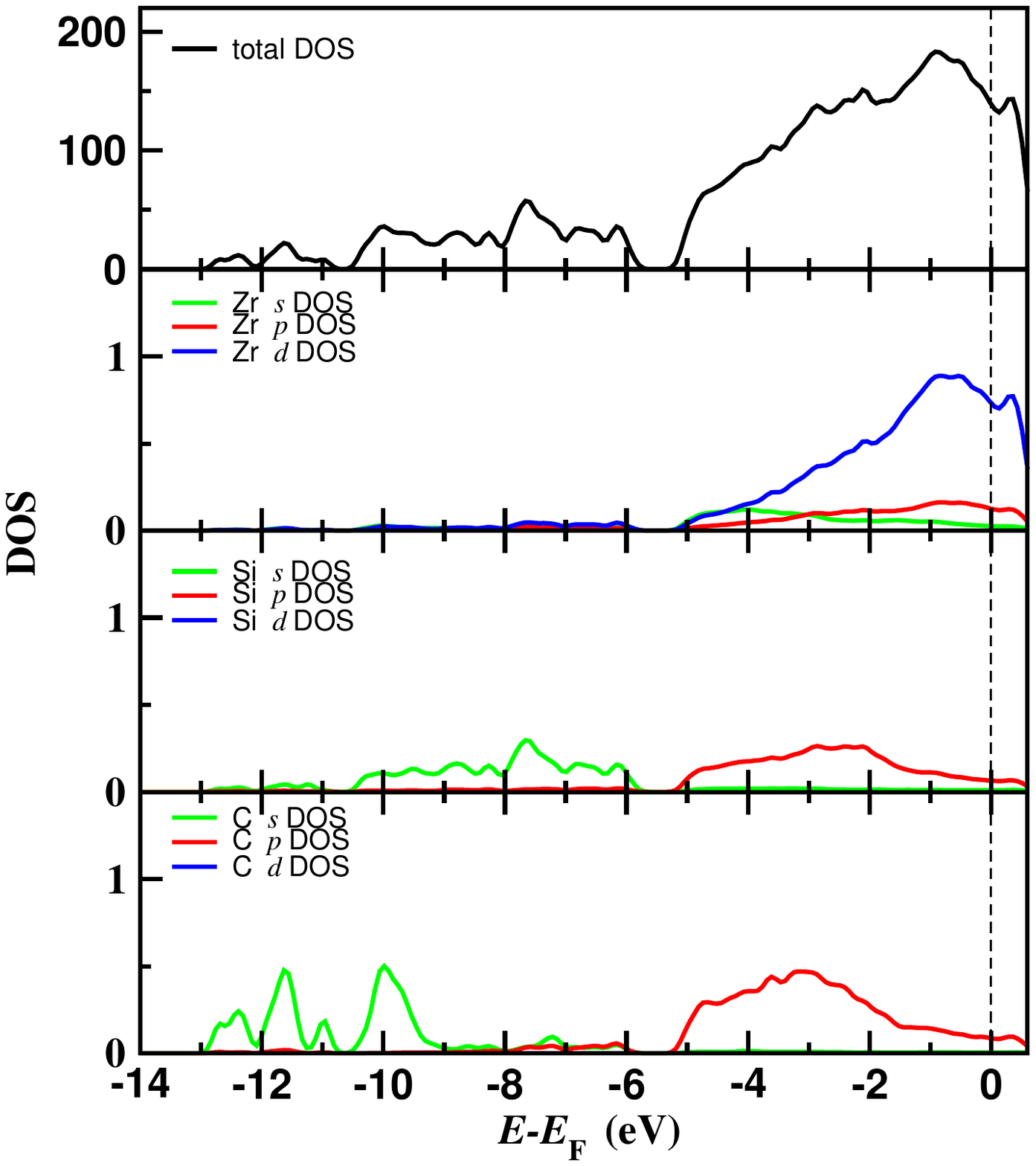}
\caption{}
\label{fig_dos}
\end{figure}

To examine the electronic structure of amorphous Zr-Si-C, we
calculated their electronic density of states (DOS).
Figure \ref{fig_dos} shows
the calculated the electronic density of states (DOS) of the
representative structure of
Zr$_{0.60}$Si$_{0.33}$C$_{0.07}$
at the equilibrium volume.
The lowest lying states at around -12 eV originate from C 2$s$ states,
and mainly Si 3$s$ orbitals contribute to states between -10.5 and -6 eV.
The large band
between -5 and 0.6 eV
is built up by Zr 4$d$, Si 3$p$, and C 2$p$ states.

We calculate a large DOS at the Fermi level 
($E_{\rm F}$)
for both compositions,
i.e. that amorphous Zr-Si-C has a metallic character.
This is in line with experimental observations:
we measured 3.20$\pm$5 $\mu \Omega$m resistivity for 
Zr$_{0.60}$Si$_{0.33}$C$_{0.07}$,
and 10.30$\pm$18 $\mu \Omega$m for Zr$_{0.31}$Si$_{0.29}$C$_{0.40}$.
The states at $E_{\rm F}$ are dominated by Zr 4$d$ orbitals,
with some
contribution from Si 3$p$ and C 2$p$ orbitals, i.e.
amorphous Zr-Si-C is conducting,
and the Zr $d$ band
has the most important role in electrical transport.
In a simple approximation, 
the density of states at the Fermi level, 
$N(E_{\rm F})$,
is directly related to the electrical conductivity.
We calculate 
$N(E_{\rm F})$=0.785 states/eV/atom for amorphous Zr$_{0.60}$Si$_{0.33}$C$_{0.07}$,
and $N(E_{\rm F})$=0.353 states/eV/atom for Zr$_{0.31}$Si$_{0.29}$C$_{0.40}$,
suggesting that Zr$_{0.60}$Si$_{0.33}$C$_{0.07}$ has a higher conductivity.
Indeed, we measured lower resistivity for Zr$_{0.60}$Si$_{0.33}$C$_{0.07}$
than for Zr$_{0.31}$Si$_{0.29}$C$_{0.40}$.
We obtain higher DOS at $E_{\rm F}$ in amorphous Zr$_{0.60}$Si$_{0.33}$C$_{0.07}$
than that
in crystalline MAX phases,
e.g. in 
Ti$_2$SiC (0.36 states/eV/atom \cite{palmquist04}),
Ti$_2$GeC (0.43 states/eV/atom \cite{hug06}),
Zr$_2$InC (0.30 states/eV/atom \cite{he09}).

\section{Conclusions}
\label{concl}

By means of the SQ method, we have generated structures 
for amorphous \\ 
Zr$_{0.31}$Si$_{0.29}$C$_{0.40}$ and Zr$_{0.60}$Si$_{0.33}$C$_{0.07}$, 
and measured the EXAFS spectra, TEM and XRD diffractograms
of samples with the same 
chemical composition.
TEM and XRD confirm that both compositions are amorphous.
We show that the calculated EXAFS spectrum 
is in 
excellent agreement with experiments,
which
serves as proof that theory
reproduces accurately the structural properties of amorphous materials. 
This demonstrates for the first time that the SQ method
provides reliable atomic coordinates of amorphous materials, even for very
complex ternary systems, with competing chemical interactions, involving
metallic and covalent bond-formation.\cite{gelatt83} 
Our findings open up a new avenue to
perform fast and accurate theoretical simulations for a wide class of amorphous
materials.

\vspace{0.6cm}

{\bf Acknowledgements}

The Swedish Research Council and 
the Swedish Foundation for Strategic Research
(SSF, via the program Technical advancement through controlled 
tribofilms) are acknowledged for financial support. O.E. also acknowledges
financial support from VR, the KAW foundation, and the ERC (247062 - ASD).
M.A. acknowledges financial support from SSF through the research program ProViking.
E.H. would like to thank for support by FONDECYT grant 1110602. 
All computations made possible by a SNAC 
allocation, on supercomputer centra UPPMAX, NSC, and HPC2N.

\clearpage

\noindent \textbf{\large{Figure captions}} \\

\noindent Figure 1: (Color online.)
X-ray diffraction patterns for Zr-Si-C films with 7 at.\% C (lower) and 40 at.\% 
carbon (upper).

\vskip 0.2cm
\noindent Figure 2: (Color online.)
TEM micrographs of Zr-Si-C film with 7 at.\% carbon.

\vskip 0.2cm
\noindent Figure 3: (Color online.)
Theoretical average energies ($E_{\rm av}$) calculated for \\
Zr$_{0.31}$Si$_{0.29}$C$_{0.40}$ (upper) and Zr$_{0.60}$Si$_{0.33}$C$_{0.07}$ (lower),
as a function of the volume, $V$.
In the insets the energy distribution of 50 stochastic configurations are shown
at $V_{\rm 0}$,
where the energies were smeared by normalized Gaussian distributions
with standard deviation of 0.01 eV.

\vskip 0.2cm
\noindent Figure 4: (Color online.)
One stochastic structure of amorphous Zr$_{0.60}$Si$_{0.33}$C$_{0.07}$,
calculated at the equilibrium volume,
having a total energy closest to the average energy of 50 
structures of the same volume (representative stucture).
Zr, Si, and C atoms are displayed by silver, yellow, and cyan, respectively.
For clear representation of the structure, bonds are not displayed in the figure.

\vskip 0.2cm
\noindent Figure 5: (Color online.)
The simulated (grey lines) and experimental (dashed lines) EXAFS
signals of amorphous Zr$_{0.31}$Si$_{0.29}$C$_{0.40}$ (in arbitrary units).
The averages over 50 individual configurations for the 200 atom case (upper) 
and over 5 configurations for the 400 atom case (lower) are shown by red lines.

\vskip 0.2cm
\noindent Figure 6: (Color online.)
The simulated (grey lines) and experimental (dashed line) EXAFS
signals of amorphous Zr$_{0.60}$Si$_{0.33}$C$_{0.07}$ (in arbitrary units).
The average of all simulated cells
are shown by the red line.

\vskip 0.2cm
\noindent Figure 7: (Color online.)
Simulated EXAFS signals (in arbitrary units) 
for some crystalline compounds compared to the
measured spectrum of amorphous \\
Zr$_{0.31}$Si$_{0.29}$C$_{0.40}$. 
The crystalline alloy is a random substitutional alloy with
the same composition as the amorphous alloy distributed on a NaCl-type crystal 
lattice with lattice constant $a$=4.698 \AA. 

\vskip 0.2cm
\noindent Figure 8: (Color online.)
Average partial radial distribution functions calculated for amorphous
Zr$_{0.31}$Si$_{0.29}$C$_{0.40}$.
The supercell size in the calculation was 200 atoms.

\vskip 0.2cm
\noindent Figure 9: (Color online.)
Average partial radial distribution functions calculated for amorphous
Zr$_{0.60}$Si$_{0.33}$C$_{0.07}$.
The supercell size in the calculation was 200 atoms.

\vskip 0.2cm
\noindent Figure 10: (Color online.)
Examples of calculated local structures around Zr (a), Si (b), and C (c) atoms
in amorphous Zr$_{0.60}$Si$_{0.33}$C$_{0.07}$.
Zr, Si, and C atoms are displayed by silver, yellow, and cyan, respectively.

\vskip 0.2cm
\noindent Figure 11: (Color online.)
Electronic density of states (in arbitrary units), calculated for 
the representative structure of amorphous Zr$_{0.60}$Si$_{0.33}$C$_{0.07}$ at the
equilibrium volume.
The top panel shows the total DOS of the 200 atom structure.
The partial $s$, $p$, and $d$ DOS curves 
are averages calculated for the different atom types.
The Fermi level ($E_{\rm F}$) is denoted by vertical dashed line.

\clearpage

\begin{table}[ht]
\begin{center}
\caption{
Average nearest neighbor distances ($d_{\rm TypeI-TypeII}$ in \AA ) calculated for
amorphous Zr$_{0.31}$Si$_{0.29}$C$_{0.40}$ and Zr$_{0.60}$Si$_{0.33}$C$_{0.07}$.
For comparison, the distances in the corresponding crystalline materials
are also listed.
}
\label{tab2}
\begin{tabular}{cccc}
\hline
& Zr$_{0.31}$Si$_{0.29}$C$_{0.40}$ & Zr$_{0.60}$Si$_{0.33}$C$_{0.07}$ & Crystalline \\
\hline
Zr-Zr &  3.22 &  3.18 & 3.179$^{a}$  \\
Zr-Si &  2.75 &  2.75 & 2.724$^{b}$, 2.723$^{c}$  \\
Zr-C  &  2.32 &  2.28 & 2.349$^{d}$  \\
Si-Si &  2.41 &  2.45 & 2.352$^{e}$  \\
Si-C  &  1.86 &  1.89 & 1.887$^{f}$  \\
C-C   &  1.41 &  1.41 & 1.545$^{g}$, 1.42$^{h}$  \\
\hline
\end{tabular}
\end{center}
\hspace{2cm}
\begin{tabular}{l}
$^{a}$: in Zr of hexagonal P6$_3$/mmc structure \cite{martienssen05} \\
$^{b}$: in ZrSi$_2$ of Cmcm structure \cite{schachner54} \\
$^{c}$: in Zr$_5$Si$_3$ of P6$_3$/mcm structure \cite{kwon90}  \\
$^{d}$: in ZrC of NaCl structure \cite{martienssen05} \\
$^{e}$: in Si of diamond structure \cite{martienssen05} \\
$^{f}$: in SiC of F$\bar{4}$3m structure \cite{martienssen05} \\
$^{g}$: in diamond \cite{martienssen05} \\
$^{h}$: in graphene \cite{martienssen05} \\
\end{tabular}
\end{table}

\clearpage

\begin{table}[ht]
\begin{center}
\caption{
Calculated average coordination numbers 
of 50 stochastic structures for
amorphous Zr$_{0.31}$Si$_{0.29}$C$_{0.40}$ and Zr$_{0.60}$Si$_{0.33}$C$_{0.07}$.
TypeI atoms (in rows) are coordinated by 
$n_{\rm TypeI,TypeII}$ number of 
TypeII atoms (in columns) on the average.
}
\label{tab3}
\begin{tabular}{c|ccc|ccc}
\hline
 & \multicolumn{3}{c|}{Zr$_{0.31}$Si$_{0.29}$C$_{0.40}$} & \multicolumn{3}{c}{Zr$_{0.60}$Si$_{0.33}$C$_{0.07}$} \\
   & Zr  &  Si & C   & Zr   &  Si & C   \\
\hline
Zr & 6.6 & 4.6 & 4.2   &   10.2 & 4.1 & 0.6   \\
Si & 4.9 & 3.6 & 1.6   &    7.4 & 1.5 & 0.2   \\
C  & 3.3 & 1.1 & 1.1   &    5.2 & 0.9 & 0.1   \\
\hline
\end{tabular}
\end{center}
\end{table}

\clearpage

\begin{table}[ht]
\begin{center}
\caption{
Calculated minimal ($m_{\rm min}$), maximal ($m_{\rm max}$) and average
($m_{\rm average}$)  number of nearest neighbors
in amorphous Zr$_{0.31}$Si$_{0.29}$C$_{0.40}$ and Zr$_{0.60}$Si$_{0.33}$C$_{0.07}$.
Numbers in parentheses give the percentages of the total number of nearest
neighbors, where the minimal and maximal values are calculated for the
corresponding individual stochastic structures.
}
\label{tab4}
\begin{tabular}{c|ccc|ccc}
\hline
& \multicolumn{3}{c|}{Zr$_{31}$Si$_{29}$C$_{40}$} & \multicolumn{3}{c}{Zr$_{60}$Si$_
{33}$C$_{7}$} \\
& $m_{\rm min}$ & $m_{\rm max}$ & $m_{\rm average}$ & $m_{\rm min}$ & $m_{\rm max}$ & $m_{\rm average}$ \\
\hline
Zr-Zr & 191 & 217 & 204.4 & 597 & 623 & 612.7 \\
& (19.4) & (22.2) & (20.7) & (48.5) & (50.7) & (49.4)  \\  \hline
Zr-Si & 258 & 305 & 285.1 & 467 & 511 & 490.2 \\
& (26.1) & (30.8) & (28.8) & (38.0) & (41.0) & (39.6)  \\  \hline
Zr-C  & 235 & 282 & 260.7 &  59 &  78 &  72.0 \\
& (23.9) & (27.9) & (26.4) &  (4.8) &  (6.3) &  (5.8)  \\  \hline
Si-Si &  90 & 120 & 104.0 &  38 &  64 &  50.2 \\
&  (9.1) & (12.3) & (10.5) &  (3.1) &  (5.2) &  (4.1)  \\  \hline
Si-C  &  77 & 108 &  90.7 &   7 &  20 &  13.2 \\
&  (7.7) & (10.8) &  (9.2) &  (0.6) &  (1.6) &  (1.1)  \\  \hline
C-C   &  34 &  57 &  44.0 &   0 &   4 &   0.9 \\
&  (3.4) &  (5.9) &  (4.5) &  (0.0) &  (0.3) &  (0.1)  \\
\hline
\end{tabular}
\end{center}
\end{table}

\clearpage

\clearpage

\clearpage

\clearpage

\clearpage

\clearpage

\clearpage

\clearpage

\clearpage

\clearpage

\clearpage

\begin{thebibliography}{10}
\bibitem{footnote1}
The Debye temperature of the random substitutional alloy was estimated to be
  between the values for ZrC and Zr$_5$Si$_3$.
\end{thebibliography}
\end{document}